\documentstyle[aps,prl,twocolumn,epsfig]{revtex}
\def\fig#1{Fig.~\ref{#1}}

\begin{document}
\draft
\title{Electronic Shell Structure of Nanoscale Superconductors}
\author{K. Tanaka and F. Marsiglio}
\address{Department of Physics, University of Alberta,
Edmonton, Alberta, Canada T6G 2J1}
\date{\today}
\twocolumn[\hsize\textwidth\columnwidth\hsize\csname@twocolumnfalse\endcsname
\maketitle
\begin{abstract}
Motivated by recent experiments on Al nanoparticles,
we have studied the effects of fixed electron number and small size 
in nanoscale superconductors, by applying the canonical 
BCS theory for the attractive Hubbard model in two and three dimensions.
A negative ``gap'' in particles with an odd number of electrons 
as observed in the experiments is obtained in our canonical scheme.
For particles with an even number of electrons,
the energy gap exhibits shell structure as a function of electron
density or system size in the weak-coupling regime:
the gap is particularly large for ``magic numbers'' of electrons for a 
given system size or of atoms for a fixed electron density.
The grand canonical BCS method essentially misses this feature.
Possible experimental methods for observing such shell effects are discussed.
\end{abstract}
]
\narrowtext

\makeatletter
\global\@specialpagefalse
\def\@oddhead{\hfill Alberta Thy 06-99}
\let\@evenhead\@oddhead
\makeatother

As the technology for fabricating ultrasmall metallic grains steadily
improves, the typical sample dimensions are approaching molecular
dimensions \cite{davidovic98,davidovic99}. The present accessibility to
samples ranging from nanoscale to bulk has renewed interest in various
features of the solid state that may or may not survive the excursion
to ultrasmall dimensions. In superconducting Al samples, for example,
the ability to distinguish even and odd numbers of electrons through
tunneling experiments \cite{tuominen92,lafarge93,ralph95}
has called into question the use of the grand canonical ensemble to
describe the electron pairing in these ultrasmall samples within a model
with equal level spacings \cite{braun98}, and within the attractive
Hubbard model \cite{tanaka99}. In these works the BCS theory of pairing was
formulated within the canonical ensemble, following the early treatment
of nuclei \cite{dietrich64}, and some rigorous ``quality control'' 
was provided by exact studies 
\cite{mastellone98,tanaka99,dukelsky99,richardson64}.

Within the attractive Hubbard model we have found two prominent features
that emerged from the canonical BCS treatment, both of which
were verified by the exact solutions. The first is the existence of
``negative gaps'' for odd electron number grains. By this we simply mean
that a tunneling bias {\it less} than the charging energy would be required
to tunnel an electron onto a grain with an odd number of electrons. The
second is the existence of what were termed ``super-even'' electron numbers,
where the tunneling bias required to tunnel an electron onto a grain with
certain even numbers of electrons would be unusually high. In this letter we
investigate these features for various bandstructures in two and three
dimensions, as might apply to Al, and briefly discuss some possible 
experiments to observe in particular the ``super-even'' effect.

We have adopted the attractive Hubbard model, whose specifics are
well known. The additional feature we include here is the possibility
of using both  periodic boundary conditions (PBC) as well as open
boundary conditions (OBC), which are more appropriate for small systems.
Either of these is accomplished through a unitary transformation to a
basis that diagonalizes the kinetic energy term. The BCS variational
calculation is then performed with a wave function containing pairs
of time-reversed states $(n,\uparrow)$ and $(n,\downarrow)$
\cite{anderson59}. The even and odd wave functions with $\nu$-pairs
are given by \cite{dietrich64}
\begin{eqnarray}
|\Psi_{2\nu}\rangle&=&c\,{1\over 2\pi i}\oint d\xi\;\xi^{-\nu-1}
\prod_{n}\,\Bigl(\,1+\xi\;g_{n}\;a_{n\uparrow}^\dagger 
a_{n\downarrow}^\dagger\,\Bigr)\,|0\rangle\;,
\label{cbcswfe}\\
|\Psi_{2\nu+1}\rangle&=&c\,{1\over 2\pi i}\oint d\xi\;\xi^{-\nu-1}\,
a_{m\sigma}^\dagger\prod_{n\neq m}\,
\Bigl(\,1+\xi\,g_{n}\,a_{n\uparrow}^\dagger 
a_{n\downarrow}^\dagger\,\Bigr)\,|0\rangle\;,
\label{cbcswfo}
\end{eqnarray}
with $N_e=2\nu$ and $N_e=2\nu+1$, respectively.
The contour integral is on any counterclockwise path that encloses the origin.
For odd $N_e$, the blocked state $m$ is chosen so that it gives the lowest
energy for a given coupling strength.
Details with PBC were given previously \cite{tanaka99}, and with OBC they
will be given elsewhere.
We calculate the ground state energy for three systems with electron number
$N_e$, $N_e+1$, and $N_e+2$, and evaluate the energy gap by the formula
$\Delta_{N_e}=(E_{N_e - 1} - 2 E_{N_e} +  E_{N_e + 1})/2$.

In the grand canonical ensemble, the number of electrons is fixed only on
average.  Thus we must solve the gap equation,
\begin{equation}
\Delta_{n}=\sum_{m}\,({\rm Re}\,V_{n n,m m})\,{\Delta_{m}\over 2 E_{m}}\;,
\label{bcs_gap}
\end{equation}
along with the number equation,
\begin{equation}
n_e \equiv {\langle N_e\rangle \over N} = 1 - {1 \over N} \sum_{n}\,
{1 \over E_{n}}\,(\tilde\epsilon_{n}  - \mu)\;,
\label{bcs_num}
\end{equation}
for gap parameters $\{\Delta_{n}\}$ and chemical potential $\mu$.
Here, $V_{n n,m m}$ is the transformed interaction, $E_{n} \equiv 
\sqrt{ (\tilde\epsilon_{n}  - \mu)^2 + \Delta_n^2}$
is the quasiparticle energy and 
$\tilde\epsilon_{n}=\epsilon_{n} + \sum_{m}\,V_{nm,nm}\,{g_m^2\over 1+g_m^2}$
is the single-particle energy modified with the Hartree term.
The gap is given by $\Delta_{0} = {\rm min}\,(E_{n})$ for a finite size system,
that is, with quantized energy levels $\{\epsilon_{n}\}$.

\begin{figure}
\begin{center}
\leavevmode
\epsfxsize=8cm
\epsfysize=10cm
\epsffile{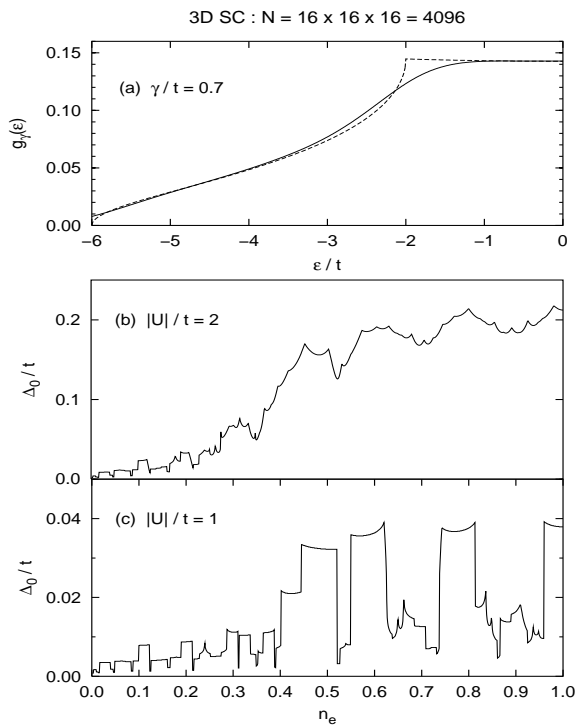}
\end{center}
\caption{(a) Density of states smoothed by Gaussian convolution with
width $\gamma=0.7 t$ for 3D $N=16^3=4096$ sites (solid curve).
The dashed curve is the bulk density of states.
(b) Energy gap obtained by the grand canonical BCS, $\Delta_{0}$/t,
as a function of electron density $n_e$ for $N=16^3=4096$ sites
and $|U|/t=2$. (c) Same as (b) but for $|U|/t=1$.
}
\label{fig1}
\end{figure}

In the weak to intermediate coupling regime,
the energy gap as a function of electron number (or electron density $n_e$)
roughly reflects the single-particle density of states (DOS).
Thus in a simple cubic (SC) lattice in three dimensions (3D),
the gap in the bulk limit is a smooth function of $n_e$ that
increases from zero at zero density \cite{remark1} to a maximum 
value at half filling.
In \fig{fig1}(a) we show the single-electron DOS for a SC lattice of
$N=16\times 16\times 16=4096$ sites with PBC (solid curve).
The result has been smoothed by convolution with
a normalized Gaussian, and it is very similar to
the bulk density of states shown by the dashed curve.
Although $N=4096$ is fairly large,
the energy gap behaves quite differently from what we expect 
for a bulk system as a function of electron density for weak coupling.
Results for the grand canonical BCS gap, $\Delta_0/t$, are
illustrated in \fig{fig1}(b) for $|U|/t=2$ and (c) for $|U|/t=1$.
For $|U|/t=2$ the overall scale of the gap $\Delta_0$ as a function of $n_e$ 
resembles $g(\epsilon)$ shown in \fig{fig1}(a).  
However, it has
many fine structures; discontinuities at small density and cusps at larger
density.  This non-smooth behaviour is a result of the discrete density 
of states, i.e., quantized energy levels $\{\epsilon_n\}$ 
and their degeneracy in a finite size system.
Such quantum structures of $\Delta_0$ turn out to be prominent
for weaker coupling strengths, as seen for $|U|/t=1$ in \fig{fig1}(c):
in this case there are discontinuities in the gap for the entire range of
density. The

\noindent discontinuities or cusps in the gap arise from finite 
level spacings, while their positions as a function of $n_e$ and 
the magnitude of the gap are determined by the degeneracy of levels, as
will be explained in detail shortly.

\begin{figure}
\begin{center}
\leavevmode
\epsfxsize=7cm
\epsfysize=8cm
\epsffile{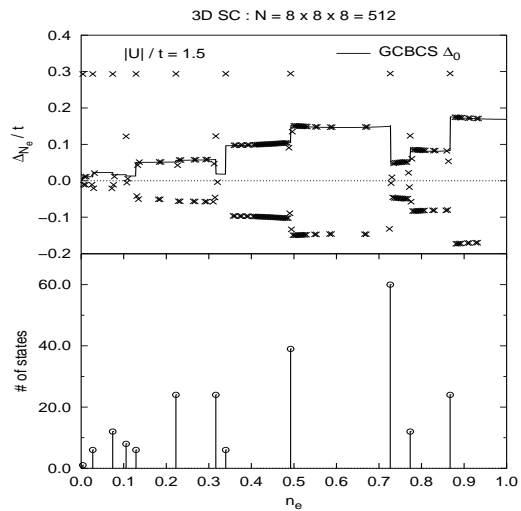}
\end{center}
\caption{Energy gap $\Delta_{N_e}$/t from the canonical BCS (crosses)
and $\Delta_{0}$/t from the grand canonical BCS (solid curve) as a function
of electron density $n_e$ for 3D $N=8^3=512$ sites and $|U|/t=1.5$
(upper frame): the canonical results are shown for only some representative
densities.
The number of single-particle states (spin degeneracy not included)
as a function of $n_e$ (lower frame): each of the discrete levels is
plotted at the density that corresponds to the closed-shell configuration
up to that level in the zero-coupling limit.
The canonical gap $\Delta_{N_e}$/t for weak coupling has jumps at these
densities as seen in the upper figure,
while the height of these jumps is determined by the energy spacing
to the next level.
}
\label{fig2}
\end{figure}

In \fig{fig2} (upper frame) the gap $\Delta_{N_e}$ obtained by the canonical
BCS is shown with crosses, along with the grand canonical gap $\Delta_0$
(solid curve), as a function of electron density $n_e$, for a nanoscale system
in weak coupling. The most obvious new feature is the ``negative gap'',
for systems with an odd number of electrons. As was already mentioned, this
result has already been observed in small Al grains \cite{ralph95}.
For the even numbered grains note that most of the results shown follow the
discontinuous, step-function-like behaviour of $\Delta_0$. However,
anomalously high values occur at densities where the grand canonical result
has discontinuities.
These anomalies follow from the analogue of shell effects for a finite 
lattice of electrons.
In the lower part of \fig{fig2}
we plot the number of
single-particle states as a function of $n_e$
for this system.
Each of the discrete levels is plotted at the density that corresponds to
electron number for filling all the levels up to that particular level
(``closed-shell'' configuration \cite{bormann92})
and the height is the degeneracy of the level without the spin factor of two.
It is clear that the densities where the canonical gap has a jump
are the ones that correspond to the closed-shell configurations.
 
In a closed-shell configuration, the occupation of levels is mainly
driven by the kinetic energy. The cost required to occupy higher
energy states exceeds the gain due to the increased interaction.
Moreover, a careful examination of the energy gain due to pairing
reveals two distinct sources, a Hartree-like term, and the explicit
BCS pairing term. The latter is small compared to the former, so
the loss in energy reduction due to less mixing of states is indeed
quite small. The Hartree-like term continues to play a role, however,
which is why the value of the gap in the closed shell configurations
is approximately equal to half the level spacing (i.e. one would
have expected an additional pairing energy).
The same physics occurs within the grand canonical ensemble \cite{tanaka99},
though the discontinuity is the best these equations can do to account
for the closed shell configurations.

\begin{figure}
\begin{center}
\leavevmode
\epsfxsize=8cm
\epsfysize=10cm
\epsffile{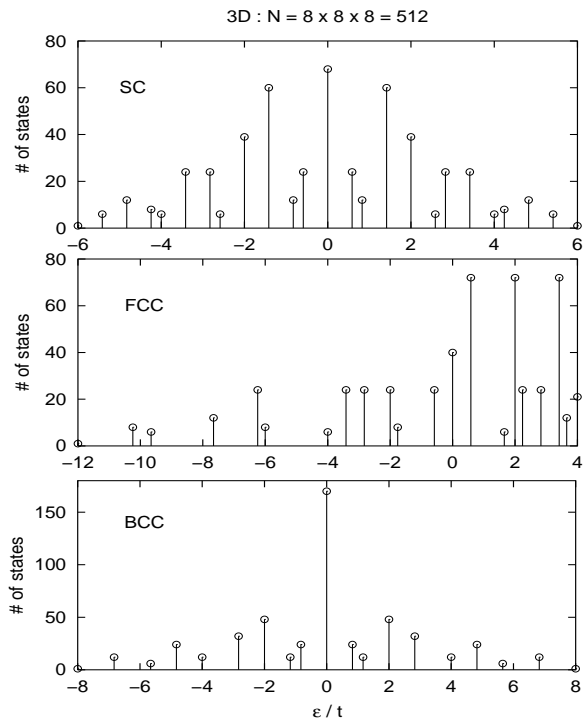}
\end{center}
\caption{The number of states (spin degeneracy not included) as a function
of single-particle energy $\epsilon$ for 3D $N=8^3=512$ sites for
different lattice structures; SC (upper frame), FCC (middle frame), and
BCC (bottom frame).  The SC and BCC level structures have
particle-hole symmetry, while FCC does not.}
\label{fig3}
\end{figure}

Different level structures result in different shell structures in the gap.
In \fig{fig3} we show the number of states for $N=8^3=512$ sites with PBC
for SC, FCC (face-centred cubic) and BCC (body-centred cubic) lattices,
as a function of single-particle energy for the entire band.
The one for SC (top frame) for negative energy is the same as shown 
in \fig{fig2}, except now it is plotted vs. energy so that the level 
spacings are clearly visible; these in turn 
determine the height of the jumps in the canonical gap.
In fact for such a small system there are only two distinct 
level spacings in SC.  This is why in the gap 
shown in \fig{fig2}, there are only two anomalously high values for the gap.
In FCC (middle frame) there is no particle-hole symmetry and the degeneracy 
is more concentrated near the top of the band.  Compared with SC,
the jumps at closed shell configurations will be more enhanced by 
the larger level spacings
and more frequent for smaller density; in addition the gap for open shells 
will be smaller (on average) because of less degeneracies.
The BCC (bottom frame) has particle-hole symmetry as in SC, but
the degeneracy is concentrated around zero energy.  As in FCC there will be
jumps more frequently at smaller densities, but near half-filling 
the gap will be continuous with the Fermi level at zero energy.

\begin{figure}
\begin{center}
\leavevmode
\epsfxsize=8cm
\epsfysize=10cm
\epsffile{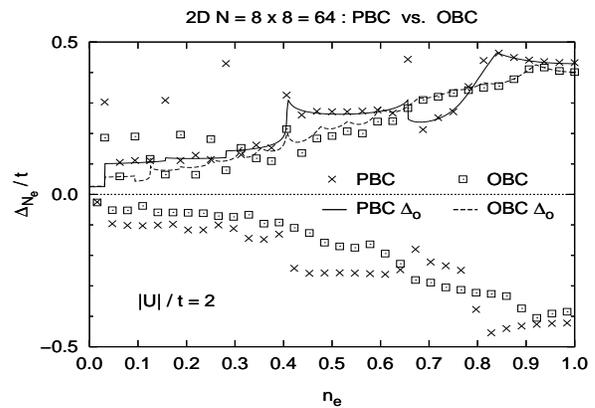}
\end{center}

\vspace{-3.5cm}
\caption{The canonical energy gap $\Delta_{N_e}$/t with PBC (crosses) and
OBC (squares) and the grand canonical $\Delta_{0}$/t with PBC (solid curve)
and OBC (dashed curve) as a function of electron density $n_e$
for 2D $N=8^2=64$ sites and $|U|/t=2$.  With OBC, due to lower symmetry and
hence less degeneracy of single-particle levels,
the canonical gap tends to have more jumps, but with less height.
}
\label{fig4}
\end{figure}

The level structure also depends on the boundary
condition.  In \fig{fig4}
we illustrate this for a small 2D system,
where we compare PBC and OBC.
In \fig{fig4} the canonical gap $\Delta_{N_e}$ is plotted 
(for all densities) with crosses (PBC) and squares (OBC), and the grand
canonical gap $\Delta_0$ is shown with solid (PBC) and dashed (OBC) curves,
for $|U|/t=2$.  
The SC lattice in 2D with PBC has a large degeneracy
(a singularity in the bulk DOS) at zero single-particle energy.  
With OBC a relatively high degeneracy remains at zero energy, 
but for nonzero energy there are more levels with less degeneracies,
because translational symmetry is absent.
For $N=64$ most of the levels are doubly degenerate, while some have no
degeneracy.  This is why in \fig{fig4} the canonical gap with OBC has
jumps more frequently (often periodic between $N_e$ multiples and non-multiples 
of four) with less height than with PBC at smaller density.
We note again that the canonical gap for even $N_e$ for open shells and
that for odd $N_e$ look symmetric about the x-axis (for both PBC and OBC).
\begin{figure}
\begin{center}
\leavevmode
\epsfxsize=8cm
\epsfysize=10cm
\epsffile{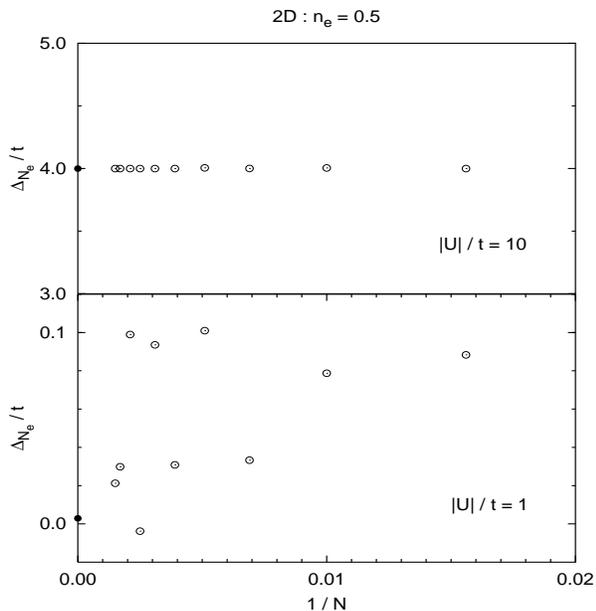}
\end{center}
\caption{The canonical gap $\Delta_{N_e}/t$ (circles) for 2D $N=N_s^2$ sites
for quarter filling $n_e=0.5$ as a function of $1/N$;
for $N_s=8$, 10, 12, 14, 16, 18, 20, 22, 24, 26 and for $|U|/t=10$ (upper
frame) and $|U|/t=1$ (lower frame).
The grand canonical gap $\Delta_{0}/t$ for $N_s=1000$ is shown on the
ordinate with a filled circle.
}
\label{fig5}
\end{figure}

We can also see the shell effects for a fixed electron density
by varying the lattice size.
In \fig{fig5} the canonical gap $\Delta_{N_e}$ 
is shown as a function of $1/N$ ($N \equiv N_s^2$) for a 2D SC lattice 
with PBC for quarter filling
$n_e=0.5$ (circles), for strong coupling (upper frame) and 
weak coupling (bottom frame).
The grand canonical gap in the bulk limit is indicated with a solid
circle on the ordinate. 
In strong coupling the gap hardly depends on
the number of sites.  In weak coupling the gap exhibits
strong size dependence as can be seen for $|U|/t=1$.
For $N_s=10$, 14, 18, 22 and 26, quarter filling corresponds to a closed-shell
configuration and this can be seen clearly for $N_s=14$, 18 and 22 as the big
jumps, which reflect the level spacing in each case.
In contrast the Fermi level is open for $N_s=8,12,16,20,24$.
Interestingly the gap for open shells also changes as a function of size 
in a non-smooth way \cite{tanakau}. By far, however, the transition to the
bulk regime is dominated by the oscillations of the magnitude of the gap
between open and closed shell configurations.

In summary, we have examined the tunneling gap for 
three dimensional ultrasmall 
superconducting grains, as a function of electron density,
coupling strength, and system size. In weak coupling, shell effects
are particularly prominent, and should be observable in very clean
grains at low temperature. An ideal experimental arrangement would
allow one to vary the electron density over a wide range. In this
way one could observe the large modulation of the gap and identify
``magic numbers'' of electrons corresponding to the electron densities
with anomalously large gap. However, in practice we anticipate that
through the use of a gate electrode, one can vary the electron density
only by a small amount (though large enough to see even/odd effects
\cite{ralph95}). Hence one will have to rely on ion-implanting a
distribution of grain sizes, and thus make use of \fig{fig5} to 
correlate gaps of different magnitude with different grain sizes. 
A systematic search should yield grain sizes whose electron number 
lies near a ``magic number'' so that tunneling a handful of electrons
(one by one) 
onto the sample controlled by a gate electrode will allow one to
observe large changes in the gap, as illustrated in \fig{fig2}.

We thank Allen Goldman, Boldizs\'ar Jank\'o, and Al Meldrum for enlightening 
discussions on possible experimental methods.
Calculations were performed on the 44-node SGI
parallel processor at the University of Alberta.
This research was supported by the Avadh Bhatia Fellowship and by the
Natural Sciences and Engineering Research Council of Canada and
the Canadian Institute for Advanced Research.


\begin{thebibliography} {999}

\bibitem{davidovic98}
D. Davidovi\'c and M. Tinkham, cond-mat/9811259.

\bibitem{davidovic99}
D. Davidovi\'c and M. Tinkham, cond-mat/9905043.

\bibitem{tuominen92}
M. T. Tuominen, J. M. Hergenrother, T.S. Tighe and M. Tinkham, Phys. Rev. Lett.
{\bf 69}, 1997 (1992).

\bibitem{lafarge93}
P. Lafarge, P. Joyez, D. Esteve, C. Urbina and M. H. Devoret, Phys. Rev. Lett.
{\bf 70}, 994 (1993).

\bibitem{ralph95}
D. C. Ralph, C. T. Black and M. Tinkham, Phys. Rev. Lett. {\bf 74}, 
3241 (1995); Phys. Rev. Lett. {\bf 76}, 688 (1996);  Phys. Rev. Lett.
{\bf 78}, 4087 (1997).

\bibitem{braun98}
F. Braun and J. von Delft, Phys. Rev. Lett. {\bf 81}, 4712 (1998).

\bibitem{tanaka99}
K. Tanaka and F. Marsiglio, Phys. Rev. B {\bf 60}, xxxx (1999).

\bibitem{dietrich64}
K. Dietrich, H. J. Mang and J. H. Pradal, Phys. Rev. {\bf 135}, B22 (1964).

\bibitem{mastellone98}
A. Mastellone, G. Falci, and R. Fazio, Phys. Rev. Lett. {\bf 80}, 4542 (1998).

\bibitem{dukelsky99}
J. Dukelsky and G. Sierra, cond-mat/9903332, cond-mat/9906166.

\bibitem{richardson64}
R.W. Richardson and N. Sherman, Nucl. Phys. {\bf 52}, 221 (1964);
R.W. Richardson, Phys. Rev. {\bf 141}, 949 (1966).

\bibitem{anderson59}
P. W. Anderson, J. Phys. Chem. Solids {\bf 11}, 26 (1959).

\bibitem{remark1}
We leave aside here the interesting issue of a minimum density required
in weak coupling and three dimensions for a bound state, and the interesting
possibility of a bound state existing in an ultrasmall system where none
exists in the bulk.

\bibitem{bormann92}
D. Bormann, T. Schneider, and M. Frick, Z. Phys. B {\bf 87}, 1 (1992).

\bibitem{tanakau}
In 1D we have found 
the canonical gap for open shells
($N$ multiples of four) and that for closed shells
($N$ non-multiples of four) are both smooth functions of $1/N$ separately
(unpublished).
The ``crossover'' behaviour of the canonical gap has been discussed
as a function of level spacing for a model with uniformly spaced levels
\cite{matveev97,mastellone98,dukelsky99}.
The crossover in the canonical BCS results was found to be rather sharp
for this model \cite{braun98}.

\bibitem{matveev97}
K. A. Matveev and A. I. Larkin, Phys. Rev. Lett. {\bf 78}, 3749 (1997).

\end{thebibliography}
\end{document}